\documentclass{article}
\usepackage{spconf,amsmath,graphicx}
\usepackage{fancyhdr}
\title{Three Dimensional MR Image Synthesis With Progressive Generative Adversarial Networks}
\name{Muzaffer Özbey$^{\star}$  \qquad Mahmut Yurt$^{\star}$  \qquad Salman Ul Hassan Dar$^{\star}$ \qquad Tolga Çukur$^{\star}$}
\address{$^{\star}$ Department of Electrical and Electronics Engineering, Bilkent University, Ankara}
\begin{document}
%
\maketitle
\begin{abstract}
Mainstream deep models for three-dimensional MRI synthesis are either cross-sectional or volumetric depending on the input. Cross-sectional models can decrease the model complexity, but they may lead to discontinuity artifacts. On the other hand, volumetric models can alleviate the discontinuity artifacts, but they might suffer from loss of spatial resolution due to increased model complexity coupled with scarce training data. To mitigate the limitations of both approaches, we propose a novel model that progressively recovers the target volume via simpler synthesis tasks across individual orientations.
\end{abstract}
\section{Introduction}
\label{sec:intro}
Current approaches for three-dimensional MRI synthesis involve either cross-sectional \cite{pgan_mri} or volumetric \cite{3dgan} mappings. Cross-sectional models can decrease model complexity by receiving as input a cross-section in a fixed orientation, e.g., axial. Yet, the synthesized images may contain discontinuity artifacts in remaning orientations since each cross-section is independently processed. On the other hand, volumetric models can alleviate these undesired artifacts by receiving an entire volume, but the synthesized images might suffer from loss of spatial resolution due to increased model complexity, especially when the training data are scarce. To address the limitations of both approaches, we propose a novel model that progressively recovers the target volume via simpler synthesis tasks across individual orientations.

\section{METHODS AND RESULTS}
\label{sec:format}
The proposed progressive approach was implemented using 3 sequential generative adversarial network (GAN) models: $\mathrm{G_A}$ that learns synthesis in the axial, $\mathrm{G_C}$ that learns synthesis in the coronal, $\mathrm{G_S}$ that learns synthesis in the sagittal orientation. In the first stage, $\mathrm{G_A}$ is trained to recover the three-dimensional MRI volume by concatenating axial cross-sections. $\mathrm{G_C}$ then takes as input the synthesized volume and it is trained to enhance the synthesis quality in coronal cross-sections. Lastly, $\mathrm{G_S}$ takes as input the output of $\mathrm{G_C}$ to further improve the synthesized volume in sagittal cross-sections. \par
\begin{figure}[htbp]
	\centering
	
	\includegraphics[scale=0.36]{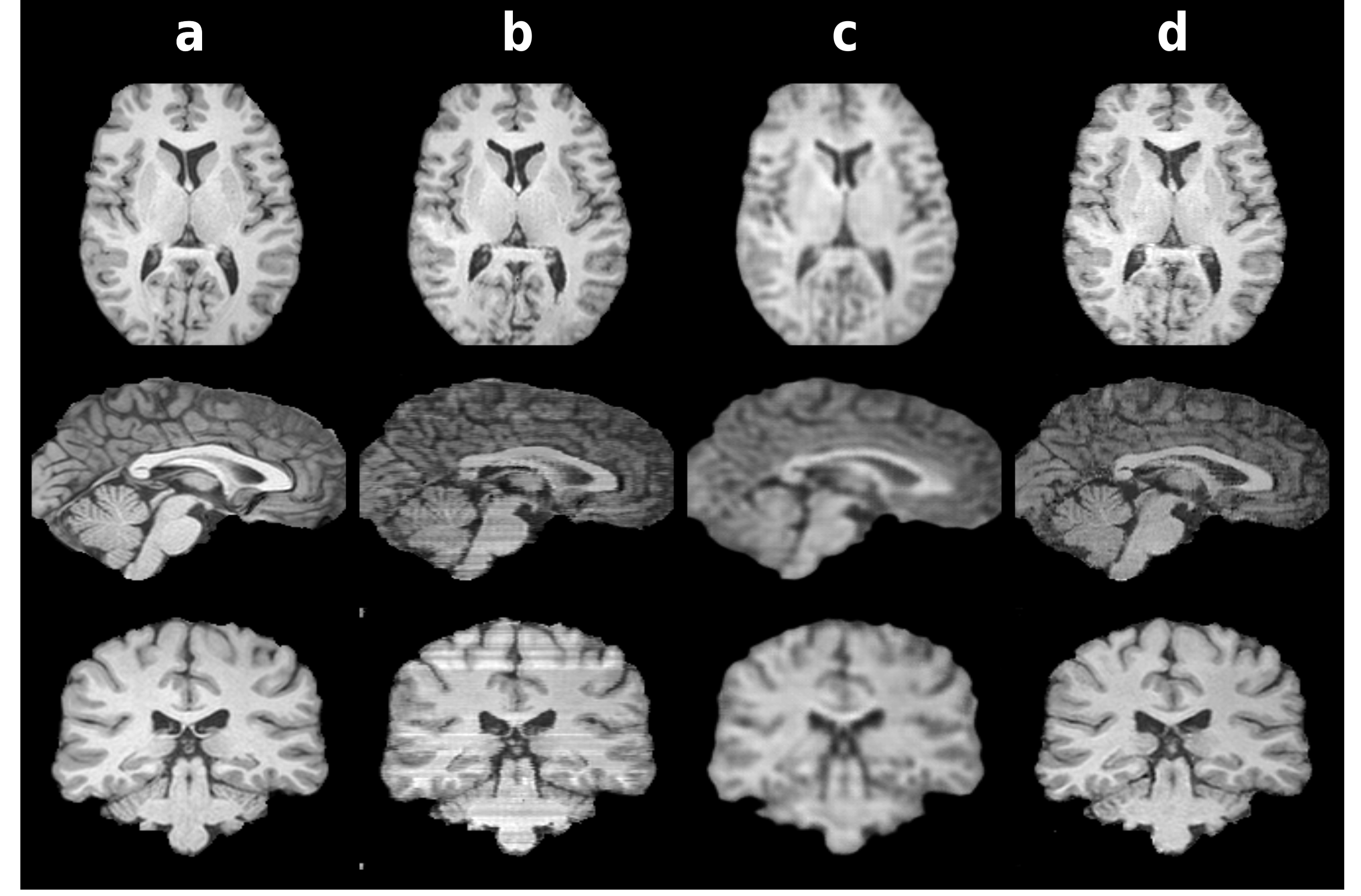}
	\caption{$T_1$-weighted volumes. a) Reference $T_1$, b) 2D-GAN, c) 3D-GAN, d) Proposed approach($\mathrm{G_A}$-$\mathrm{G_C}$-$\mathrm{G_S}$).}
	\label{syn_res}
\end{figure}

 
Comprehensive evaluations were performed on the IXI dataset for $T_1$ synthesis from $PD$ and $T_2$ with 35 training, 5 validation and 10 test subjects. The proposed approach was compared with cross-sectional GAN (2D-GAN) \cite{pgan_mri} and volumetric GAN (3D-GAN) \cite{3dgan}. On average, the proposed approach achieves 0.87 dB higher PSNR and 36.67\% lower FID compared to 2D-GAN and 2.44 dB higher PSNR and 42.24\% lower FID compared to 3D-GAN. Representative results in Fig. \ref{syn_res} also demonstrate that the proposed approach maintains high spatial resolution in synthesized volumes without introducing discontinuity artifacts.

\bibliographystyle{IEEEbib}
\bibliography{refs}

\begin{thebibliography}{1}

\bibitem{pgan_mri}
Dar et. al.,
\newblock ``Image synthesis in multi-contrast {MRI} with conditional generative
  adversarial networks,''
\newblock {\em IEEE Trans. Med. Im}, pp. 2375--2388, 2019.

\bibitem{3dgan}
Yu~et. al.,
\newblock ``{3D} c{GAN} based cross-modality {MR} image synthesis for brain
  tumor segmentation,''
\newblock in {\em 2018 IEEE 15th Int. Sym. on Bio. Im (ISBI 2018)}. IEEE, 2018,
  pp. 626--630.

\end{thebibliography}

\end{document}